\documentclass[aps,prl,twocolumn,groupedaddress,showpacs]{revtex4}

\usepackage{graphicx}
\usepackage{dcolumn}
\usepackage{bm}

\begin{document}


\title{Universality of Bias- and Temperature-induced Dephasing
 in Ballistic Electronic Interferometers}

\author{Y. Yamauchi$^1$, M. Hashisaka$^1$, S. Nakamura$^1$,
 K. Chida$^1$, S. Kasai$^1$, T. Ono$^1$, R. Leturcq$^2$, K. Ensslin$^3$,
 D. C. Driscoll$^4$, A. C. Gossard$^4$, and K. Kobayashi$^1$}


\affiliation{$^1$Institute for Chemical Research, Kyoto University, Uij,
Kyoto 611-0011, Japan}

\affiliation{$^2$IEMN - CNRS, Cit{\'e} Scientifique, Avenue Poincar{\'e}
BP 60069, Villeneuve d'Ascq, France} \affiliation{$^3$Solid State
Physics Laboratory, ETH Z\"{u}rich, CH-8093 Z\"{u}rich, Switzerland}

\affiliation{$^4$Materials Department, University of California, Santa
 Barbara, California 93106, USA}

\date{\today}

\begin{abstract}
We performed a transport measurement in a ballistic Aharonov-Bohm ring
and a Fabry-P{\'e}rot-type interferometer. In both cases we found that
the interference signal is reversed at a certain bias voltage and that
the visibility decays exponentially as a function of temperature, being
in a strong analogy with recent reports on the electronic Mach-Zehnder
interferometers.  By analyzing the data including those in the previous
works, the energy scales that characterize the dephasing are found to be
dominantly dependent on the interferometer size, implying the presence
of a universal behavior in ballistic interferometers in both linear and
non-linear transport regimes.
\end{abstract}



\maketitle

Electron interference has been the central issue in mesoscopic physics
since 1980's, which has been promoting our understanding on the phase
and coherence of electrons in solids~\cite{Imry1997}. Recently, an
electronic Mach-Zehnder interferometer (MZI)~\cite{JiNature2003} is
invoking considerable interest for its possibility to create electron
entanglement by two-particle interference~\cite{SamuelssonPRL2004,
NederNature2007}.  This MZI relies on the edge channel transport as it
requires chirality of the electron beam, which is also an advantage as
the coherence length of the edge electrons is expected to be
long. Several experimental works on MZI~\cite{JiNature2003,
LitvinPRB2007, RoulleauPRL2008, LitvinPRB2008}, however, reported that
the interference visibility, a measure of coherence, decreases
exponentially as the temperature increases.  Furthermore, the
experimentalists found~\cite{NederPRL2006,
RoulleauPRB2007,RoulleauPRL2008,LitvinPRB2008} the so-called ``lobe
structure'', an unexpected phase reversal in the visibility at a certain
bias voltage.

Several theoretical attempts~\cite{ChalkerPRB2007, LevkivskyiPRB2008,
YounPRL2008, NederPRL2008} have tried to explain the observed small
energy scales that characterize the lobe structure as well as its
drastic temperature dependence. The Coulomb interaction was proposed to
be responsible for the observed energy scale proportional to the inverse
of the size of MZI~\cite{LevkivskyiPRB2008, YounPRL2008,
NederPRL2008}. It is, however, still unclear whether or not the above
observations are specific to MZI. Indeed, some
theories~\cite{YounPRL2008, NederPRL2008} suggest that similar effects
may occur in other ballistic interferometers. To investigate this
possibility constitutes the central motivation of the present
experimental work.

Here we report the conductance measurements for an Aharonov-Bohm ring
(ABR) around zero magnetic fields and for a Fabry-P{\'e}rot-type
interferometer (FPI) in the integer quantum Hall (IQH) regime. In both
cases, we observed the lobe-like structure and the exponential decay of
the visibility with temperature, akin to those in MZI. By analyzing the
present data as well as those obtained in previous works, the
characteristic energy scales of the two effects are found to dominantly
depend on the interferometer size.

\begin{figure}[tbp]
\begin{center}
\center \includegraphics[width=.9\linewidth]{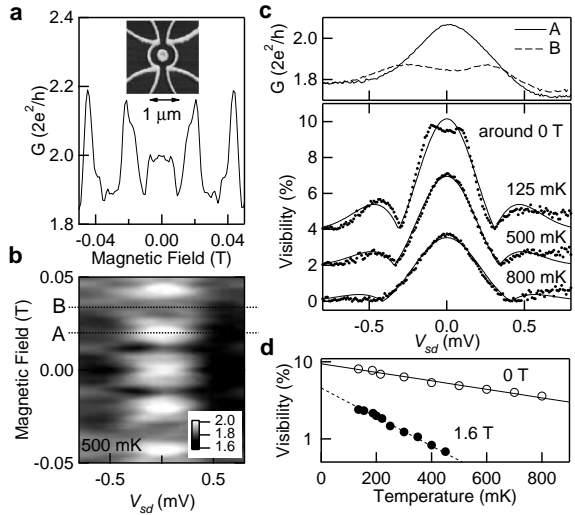} \caption{
(a) Conductance of the ABR at 125~mK as a function of $B$. Inset:
AFM image of the sample, white lines are oxidized area, under which the
2DEG is depleted~\cite{LeturcqPhysicaE2006}.  (b) Image plot of the
conductance as a function of $V_{sd}$ and $B$.  The color scale is shown
in units of $2e{^2}/h$. (c) The upper panel shows the line profiles at
the fixed magnetic field indicated by ``A'' and ``B'' in (b).  The lower
panel shows the corresponding visibility as a function of $V_{sd}$ at $T
= 125$, 500, and 800~mK.  The solid lines are the result of the fitting
to Eqn.~(1). They are vertical offset upward for clarity. (d) Electron
temperature dependence of the zero-bias visibility for 0~T and 1.6~T
(applied perpendicular to 2DEG) plotted in a semi-logarithmic scale.
The solid and dashed lines show the exponential decay function of $T$.}
\label{Fig1}
\end{center}
\end{figure}

Figure~1a shows the atomic force microscope (AFM) image of the ABR
fabricated by local oxidation using an AFM~\cite{HeldAPL1998} on a
GaAs/AlGaAs heterostructure two-dimensional electron gas (2DEG) (the
electron density $3.7 \times 10^{11}$~cm$^{-2}$ and the mobility $2.7
\times 10^5$~cm$^2$/Vs)~\cite{LeturcqPhysicaE2006}.  Measurements were
performed around zero magnetic field and in the IQH regime. The
FPI~\cite{vanWeesPRL1989,BirdPRB1994} was patterned by using 
electron beam lithography technique on an GaAs/AlGaAs heterostructure
2DEG (the electron density of $2.4 \times 10^{11}$~cm$^{-2}$ and the
mobility $2.7 \times 10^5$~cm$^2$/Vs). The scanning electron micrograph
(SEM) image of the sample is shown in the inset of Fig.~2a. We tuned the
magnetic field ($B$) and the gate voltages so as to make a small
interferometer in the IQH regime~\cite{vanWeesPRL1989,BirdPRB1994}.  For
both samples, the two-terminal conductance as well as the differential
conductance at finite dc source drain voltages ($V_{sd}$) were measured
by using lock-in techniques at 37~Hz with 5~$\mu$V ac bias
voltage. The samples were placed in a dilution refrigerator and the
electron temperature is deduced from thermal noise
measurement~\cite{HashisakaPRB2008} with the lowest electron temperature
($T$) of 125~mK.

First we show the result of the ABR below 50~mT. At zero bias voltage, a
clear AB oscillation with a visibility (the ratio of the oscillatory to
the non-oscillatory conductance components) larger than 8~\% at 125~mK
is observed (Fig.~1a).  The period of the AB oscillation is 25~mT, being
consistent with the lithographic ring geometry with a radius of
230~nm~\cite{LeturcqPhysicaE2006}.  Figure 1b presents the image plot of
the conductance as a function of $B$ and $V_{sd}$.  Importantly, a phase
reversal of the oscillation is found to occur around
$\left|V_{sd}\right|\sim 0.35$~mV. In the upper panel of Fig.~1c the
cross sections of Fig.~1b are shown at the lines denoted by ``A'' and
``B'', which correspond to the peak and dip of the AB oscillation at
$V_{sd}$ = 0~V, respectively. Because of the phase reversal of the
oscillation, the resultant visibility ($\nu$) presents a structure akin
to the ``lobe structure'' known in MZI (see the bottom panel of
Fig.~1c)~\cite{NederPRL2006,RoulleauPRB2007,RoulleauPRL2008,LitvinPRB2008}.
A similar phase reversal of the oscillation in the ABR was reported
before~\cite{YacobyPRB1996}, while no clear explanation has been given
so far.

Figure 1c shows the visibility obtained at $T = 125$, 500, and
800~mK. The characteristic energy scale of the structure is estimated by
using the following empirical equation as a function of
$V_{sd}$~\cite{LitvinPRB2008}:
\begin{equation}
\nu = \nu{_0}\left|\cos\left(\frac{{\pi}eV_{sd}}{\epsilon{_L}}\right)
\right|\exp\left(\frac{-(eV_{sd}){^2}}{2\epsilon{_0}{^2}}\right).
\end{equation}
This is the product of the absolute value of a cosine curve, with period
$\epsilon{_L}$, which characterizes the energy scale of the bias-induced
phase reversal, and a Gaussian envelope with characteristic width
$\epsilon{_0}$, which corresponds to the bias dephasing factor.  $\nu_0$
is the zero-bias visibility.  From the numerical fitting, $\epsilon{_0}$
and $\epsilon{_L}$ at 125~mK are $0.30$~meV and $0.62$~meV,
respectively~\cite{CommentDip}. While the visibility rapidly decreases
when the temperature changes from 125 to 800~mK, $\epsilon{_L}$ and
$\epsilon{_0}$ are almost independent of temperature (Fig.~1c).

Although the characteristic energy scales are more than 10 times larger,
these results look similar to those obtained for
MZI~\cite{NederPRL2006, RoulleauPRB2007, RoulleauPRL2008,
LitvinPRB2008}. It is, however, important to examine here whether or not
the phase reversal is explained by the magneto-electric AB
effect~\cite{vanderWielPRB2003} in the single particle picture.  If the
electrostatic potential difference between the two interfering paths in
the ABR were responsible for the present observation, the bias
dependence of the visibility would be determined by the term $\cos
(eV_{sd}t_0/\hbar)$~\cite{vanderWielPRB2003, NederPRL2006}. Here $t_0$
is the time for the electrons staying in the paths with different
electrostatic potentials. In the present case, the condition to obtain
$\epsilon{_L} = 0.62$~meV yields $t_0 \sim 3$~ps and therefore the
electron traveling length, which would correspond to a path difference,
is $0.7~\mu$m as the Fermi velocity of the present 2DEG is $2.5 \times
10^5$~m/s.  However, such a path difference is unfeasible in the present
symmetric geometry.  Thus, the present lobe-like structure cannot simply
be explained by the magneto-electric AB effect, which is also the case
in MZI~\cite{NederPRL2006}.

The analogy between the present result and those in MZI also lies in
the fact that the zero-bias visibility has a clear exponential
dependence on the electron temperature as shown in Fig.~1d. By fitting
the temperature dependence to the function $\nu = \nu{_0}\exp(-T/T{_0})$
we obtain $T_0 = 870 \pm 30$~mK. The exponential decrease of the
visibility with $T$ is consistent with previous reports for the ABR
around $B=0$~T~\cite{HansenPRB2001, KobayashiJPSJ2002}. As the thermal
broadening effect whose characteristic time scale is given by $\hbar/k_B
T$ predicts $T_0 = 2.5$~K for the size and the Fermi velocity of the
ABR~\cite{KobayashiJPSJ2002}, the present dephasing is not explained by
the thermal effect alone.

\begin{figure}[tbp]
\begin{center}
\center \includegraphics[width=.9\linewidth]{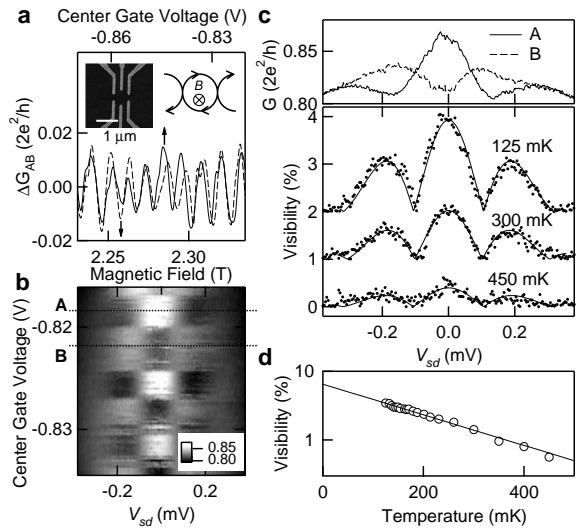} \caption{
(a) The oscillating components in the conductance of the FPI
($\Delta G_{AB}$) as a function of $B$ (dashed line) and the center gate
voltage (solid line).  The insets show the SEM image of the sample in
the left and the schematic drawing of the edge channels to construct the
FPI in the right. (b) Image plot of the conductance as a function of
$V_{sd}$ and the center gate voltage.  The color scale is shown in units
of $2e{^2}/h$. (c) The upper panel shows the line profiles at the fixed
gate voltages indicated by ``A'' and ``B'' in (b).  The lower panel
shows the corresponding visibility as a function of $V_{sd}$ at $T=125$,
300, and 450~mK.  The solid lines are the result of the fitting to
Eqn.~(1).  They are vertically offset upward for clarity. (d) Electron
temperature dependence of the zero-bias visibility is plotted in a
semi-logarithmic scale.  The line shows the exponential decay function
of $T$.} \label{Fig2}
\end{center}
\end{figure}

Next we discuss that a perfectly similar phenomenon occurs in the
FPI. The conductance measurements for the FPI were performed in the IQH
regime with a filling factor of 4 at $B \sim 2.26$~T. The gate voltages
of the left and right pairs of the metallic gates in the SEM picture in
Fig.~2a were fixed to -1.1~V and -1.0~V, respectively, so as to tune the
conductance of each point contact around $1.6e^2/h$. The pair of the
center gates were chosen to be around -0.8~V to deplete the 2DEG
underneath. By doing this, the one-dimensional state consisting of edge
channels is formed along the area defined by the six gates, which is a
small interferometer (see the upper right inset of
Fig.~2a)~\cite{vanWeesPRL1989,BirdPRB1994}. Figure~2a represents the
oscillating component in the conductance through the FPI ($\Delta
G_{AB}$) as a function of $B$ or the center gate voltage.  The AB period
is 11.8~mT, indicating that the radius of the interferometer is around
350~nm~\cite{vanWeesPRL1989} being consistent with the lithographic size
of the device.

Figure 2b shows the image plot of the conductance as a function of the
gate voltage and $V_{sd}$. As in Fig.~1a, the present result shows the
phase reversal of the oscillation at a finite bias voltage. From the
cross section at the lines denoted ``A'' and ``B'' (Fig.~2b), the
visibility is obtained as presented in Fig.~2c, which again looks
similar to the lobe structure. Such a lobe-like structure is also
obtained from the differential conductance data as a function of
$V_{sd}$ and $B$ instead of the gate voltage. The result of the
temperature dependence of the lobe structure and the visibility at zero
bias voltage between 125~mK and 450~mK are shown in Figs.~2c and 2d,
respectively. Just as we did above, $\epsilon{_0}$ and $\epsilon{_L}$
are obtained to be 0.18~meV and 0.21~meV, respectively. $T{_0} \sim$
200~mK, which again cannot be explained by the thermal broadening
effect.

Importantly, a completely similar phenomenon was observed in the ABR at
$B \sim$ 1.6~T (filling factor 10). In this case, we obtain
$\epsilon{_0} = 0.25$~meV and $\epsilon{_L} =0.26$~meV with an AB period
of 17~mT. The visibility also decreases exponentially with temperature
with $T{_0} = 240$~mK as shown in Fig.~1d.  The different values of
$\epsilon_0$ and $\epsilon_L$ compared to those at $B=0$~T are due to a
different interfering path geometry as indicated by the different AB
period. However, it is noteworthy that $T{_0}$ is much reduced from the
zero-field value, as will be discussed below.

Recently it was theoretically~\cite{YounPRL2008, NederPRL2008} proposed
that the lobe structure in MZI occurs through the Coulomb
interaction; When the number of excess electrons present in MZI
becomes two in the non-linear regime, the phases of the electrons are
mixed due to the Coulomb interaction, resulting in the rapidly
decreasing visibility with the phase reversal. Following their idea, a
simple estimation of the energy scale for the present lobe-like
structures can be performed for the ABR and the FPI as follows; The
above effect occurs at $V_{sd} = V_{0}$ when the electron traveling time
in one arm $L/v{_f}$ is almost twice the injection period $e/G|V_{0}|$
for the system with the conductance $G$. Here, $L$ is the arm length
calculated by the oscillation period and $v{_f}$ is the Fermi velocity
($2.5 \times 10^5$~m/s for the ABR at $B=0$~T) and the edge velocity for
the FPI (typically $3 \times 10^4$~m/s), respectively.  The value
$eV_{0}$ to characterize this effect is obtained to be 0.74~meV for the
ABR at $B=0$~T ($L = 0.72~\mu$m and $G \sim 4e^2/h$) and to be 0.15~meV
for the FPI ($L = 1.1~\mu$m and $G \sim 1.6e^2/h$). As these values are
comparable to the observed $\epsilon{_0}$ or $\epsilon{_L}$ in each
case, given the simplicity of the discussion, the dephasing due to the
correlation between the excess electrons satisfactorily explains what
happens in the non-linear electronic interferometer.

\begin{figure}[tbp]
\begin{center}
\center \includegraphics[width=0.99\linewidth]{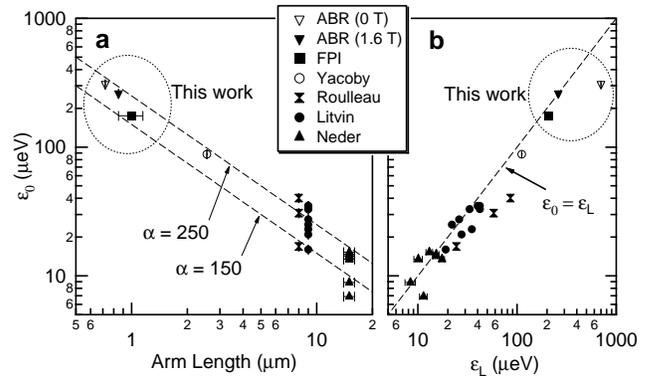} \caption{ (a)
The energy scale ($\epsilon{_0}$) of the lobe structure is plotted as a
function of the arm length ($L$) by using the data for the present ABR
at 0~T and 1.6~T, the present FPI, the ABR at 0~T reported by Yacoby
{\it et al.}~\cite{YacobyPRB1996}, and MZI by Roulleau {\it et
al.}~\cite{RoulleauPRB2007,RoulleauPRL2008}, Litvin {\it et
al.}~\cite{LitvinPRB2008}, and Neder {\it et al.}~\cite{NederPRL2006}
Open points represent the data at zero magnetic field, and filled points
in the IQH regime.  The arm length of MZI by Neder {\it et
al.}~\cite{NederPRL2006} is estimated from the SEM image of their
sample. Different points with the same mark correspond to data obtained
at different conditions of the magnetic field, the opening of the beam
splitters, and so on. The two dashed lines indicate $\epsilon{_0} =
\alpha/L$ with $\alpha = 150$ and 250~$\mu$eV$\cdot\mu$m.  (b) The plot
of $\epsilon{_L}$ vs $\epsilon{_0}$ shows that the two values are almost
in the same range for a wide parameter range. } \label{Fig3}
\end{center}
\end{figure}

Having proven that the above discussion essentially captures the
dephasing in the ABR and FPI in the non-linear regime, it is instructive
to compare the present results with previous reports on ballistic
interferometers to check the universality. The dependence of
$\epsilon{_0}$ on the length of the arm of the interferometer ($L$) (the
averaged length between the two arms) is compiled in Fig.~3a.  We
plotted the data obtained for the present ABR at 0~T and 1.6~T, the
present FPI, the ABR at 0~T~\cite{YacobyPRB1996}, and
MZI~\cite{RoulleauPRB2007,RoulleauPRL2008,LitvinPRB2008,NederPRL2006}.
Interestingly, this graph shows that $\epsilon_0$ scales inversely
proportional to the arm length, $\epsilon_0 = \alpha / L$, where
$\alpha$ falls between 150 and 250~$\mu$eV$\cdot\mu$m as shown in
Fig.~3a.  Fig.~3b shows the relation between $\epsilon{_0}$ and
$\epsilon{_L}$, showing that $\epsilon{_0} \sim \epsilon{_L}$ holds for
a wide range of the interferometer sizes. $\alpha$ is determined in
principle in the same way as above by taking into account the Fermi (or
group) velocity and the linear conductance.  As theory
suggests~\cite{YounPRL2008, NederPRL2008}, the variation of $\alpha$ in
Fig.~3a and the slight deviation between $\epsilon{_0}$ and
$\epsilon{_L}$ may be explained by including the effective Coulomb
interaction in the interferometers with the realistic geometry.  Also it
is experimentally shown that $\epsilon{_0}$ and $\epsilon{_L}$ can be
controlled by the opening of the beam splitter~\cite{RoulleauPRB2007} or
the magnetic field~\cite{LitvinPRB2008,NederPRL2006}, suggesting that
the arm length is not the only parameter to affect the lobe
structure. However, Fig.~3a clearly demonstrates that the relevant
energy scale to describe the lobe structure in several kinds of
interferometers is primarily determined by the size, which is the
central result of the present experimental work.

We also found that $T{_0}$, which characterizes the dephasing in the
linear regime, strongly depends on $L$ as shown in Fig.~4, where we
compile the $T_0$ values obtained for the present ABR at 0~T and 1.6~T,
the present FPI, the FPI~\cite{BirdPRB1994}, the
ABR~\cite{HansenPRB2001,KobayashiJPSJ2002,CassePRB2000} and
MZI~\cite{RoulleauPRB2007,RoulleauPRL2008,LitvinPRB2008}. The plots
indicate another universal behavior $T_0 = \beta/L$ with $\beta =
200$~mK$\cdot\mu$m for the edge transport regime~\cite{RatioAlphaBeta}.
However, the data for the ABR at zero magnetic field strongly deviate
from this line. This suggests that the difference of the group velocity
between both is important and the coherence time but not the coherence
length matters.  The present notable difference between the zero-field
state and the edge state may suggest that the environmental property in
the vicinity of the interferometer is responsible for the
temperature-induced dephasing~\cite{KobayashiJPSJ2002,
SeeligPRB2003,RoulleauPRL2008_2}.

\begin{figure}[tbp]
\begin{center}
\center \includegraphics[width=0.8\linewidth]{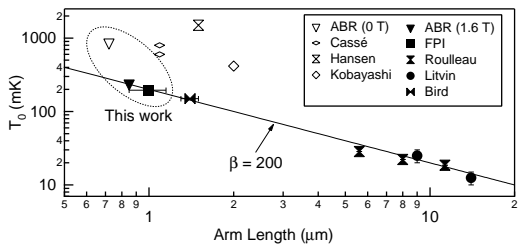} \caption{The
dependence of $T_0$ on the arm length ($L$) of the interferometers is
plotted by using the data obtained for the present ABR at 0~T and 1.6~T,
the present FPI, the FPI by Bird {\it et al.}~\cite{BirdPRB1994}, the
ABR by Cass{\'e} {\it et al.}~\cite{CassePRB2000}, Hansen {\it et
al.}~\cite{HansenPRB2001} and Kobayashi {\it et
al.}~\cite{KobayashiJPSJ2002} and MZI by Roulleau {\it et
al.}~\cite{RoulleauPRB2007,RoulleauPRL2008}, Litvin {\it et
al.}~\cite{LitvinPRB2008}. Filled points were those obtained under IQH
regime, while opened points were at zero magnetic field.  } \label{Fig4}
\end{center}
\end{figure}

To conclude, we found the ``lobe structure'' in the ballistic ABR and
FPI, being analogous to those observed in MZI's.  The characteristic
energy scales are semi-quantitatively explained as due to the Coulomb
interaction between the excess electrons in the interferometer.  By
compiling the energy scales for the lobe structure and the temperature
dependence of the visibility, they are found to be proportional to the
inverse of the interferometer size. The presence of such universality in
dephasing, which is unveiled through the measurement on the present
small electron interferometers, will shed new light on the coherence in
ballistic systems.

We appreciate fruitful comments from H. W. Lee and A. Helzel.  This work
is supported by KAKENHI, Yamada Science Foundation, and Matsuo Science
Foundation.


\begin{thebibliography}{99}

\bibitem{Imry1997} Y. Imry, Introduction to Mesoscopic Physics(Oxford
University Press, Oxford, 1997).

\bibitem{JiNature2003} Y. Ji {\it et al.}, Nature (London) {\bf 422,}
415 (2003).


\bibitem{SamuelssonPRL2004} P.~Samuelsson, E.~V.~Sukhorukov, and
M.~B{\"u}ttiker, Phys.  Rev. Lett. {\bf 92,} 026805 (2004).

\bibitem{NederNature2007} I. Neder {\it et al.}, Nature (London) {\bf
448,} 333 (2007).


\bibitem{RoulleauPRL2008} P. Roulleau {\it et al.},
Phys. Rev. Lett. {\bf 100,} 126802 (2008).


\bibitem{LitvinPRB2007} L.~V.~Litvin, H.-P.~Tranitz, W.~Wegscheider, and
C.~Strunk, Phys. Rev. B {\bf 75,} 033315 (2007).

\bibitem{LitvinPRB2008} L. V. Litvin {\it et al.}, Phys. Rev. B {\bf
78,} 075303 (2008).


\bibitem{NederPRL2006} I. Neder {\it et al.}, Phys. Rev. Lett. {\bf 96,}
016804 (2006).


\bibitem{RoulleauPRB2007} P. Roulleau {\it et al.}, Phys. Rev. B
{\bf 76,} 161309 (2007).



\bibitem{ChalkerPRB2007} J. T. Chalker, Y. Gefen, and M. Y. Veillette, Phys. Rev. B
{\bf 76,} 085320 (2007) and references therein.

\bibitem{LevkivskyiPRB2008} I.~P.~Levkivskyi and E.~V.~Sukhorukov, 
Phys. Rev. B {\bf 78,} 045322 (2008).

\bibitem{YounPRL2008} S. -C. Youn, H.-W. Lee, and H. -S. Sim 
Phys. Rev. Lett. {\bf 100,} 196807 (2008).

\bibitem{NederPRL2008} I. Neder and E. Ginossar 
Phys. Rev. Lett. {\bf 100,} 196806 (2008).

\bibitem{HeldAPL1998} R. Held {\it et al.}, Appl. Phys. Lett. {\bf 73,}
262 (1998).


\bibitem{LeturcqPhysicaE2006} R. Leturcq {\it et al.}, Physica E {\bf
35,} 327 (2006).


\bibitem{vanWeesPRL1989} B. J. van Wees, {\it et al.}, Phys. Rev. Lett
{\bf 62,} 2523 (1989).


\bibitem{BirdPRB1994} J.~P.~Bird {\it et al.},Phys. Rev. B {\bf 50,}
14983 (1994).


\bibitem{HashisakaPRB2008} M. Hashisaka {\it et al.}, Phys. Rev. B {\bf
78,} 241303(R)(2008).


\bibitem{YacobyPRB1996} A. Yacoby, R. Schuster, and M. Heiblum,
Phys. Rev. B {\bf 53,} 9583 (1996).

\bibitem{CommentDip} The small dip around $V_{sd} = 0$~V in the 125~mK
data is due the mesoscopic fluctuation inside the ABR, which does not
affect the present discussion on the characteristic energy scales.
The dip disappears above 400~mK.

\bibitem{vanderWielPRB2003} W. G. van der Wiel {\it et al.},
Phys. Rev. B {\bf 67,} 033307 (2003).


\bibitem{HansenPRB2001} A. E. Hansen {\it et al.}, Phys. Rev. B {\bf
64,} 045327 (2001).


\bibitem{KobayashiJPSJ2002} K. Kobayashi, H. Aikawa, S. Katsumoto, and
Y. Iye, J. Phys. Soc. Jpn 71, L2094 (2002).

\bibitem{CassePRB2000} M.Cass{\'e} {\it et al.,} Phys. Rev. B {\bf 62,} 2624 (2000).


\bibitem{SeeligPRB2003} G. Seelig, S. Pilgram, A. N. Jordan, and
M. B{\"u}ttiker, Phys. Rev. B {\bf 68,} 161310R (2003).

\bibitem{RoulleauPRL2008_2} P. Roulleau {\it et al.},
Phys. Rev. Lett. {\bf 101,} 186803 (2008).


\bibitem{RatioAlphaBeta} Recently it was theoretically discussed that
there is a universal relation between $\alpha$ and
$\beta$~\cite{LevkivskyiPRB2008}.

\end{thebibliography}
\end{document}